\documentclass{article}
\usepackage{setspace}
\usepackage{graphicx}
\usepackage{amsmath,amssymb}
\begin{document}
\title{Local mathematics and scaling field: effects on local physics and on cosmology}
\author{Paul Benioff \\Physics division, Argonne national Laboratory\\Argonne, Illinois\\
  email: pbenioff@anl.gov}
\date{}
\maketitle
\tableofcontents
\large

\section{Origin of this work}

The origin of this paper starts with the observation  by Yang Mills \cite{YangM}  that what state represents a proton in isospin space at one location does not determine what state represents a proton in isospin space  at another location. This is accounted for by the presence of a unitary gauge transformation operator, $U(y,x)$, between vector spaces at different locations.  This operator defines the notion of same states for vector spaces at different locations. If $\psi$ is a state in a vector space at $x$ then $U(y,x)\psi$  is the same state in the vector space at $y$.\footnote{The unitary gauge transformation can depend on the path from $x$ to $y.$}

            Vector spaces include scalar fields in their axiomatic description,
           These appear as norms, closure under vector scalar multiplication, etc. This leads to a conflict: local vector spaces and global scalar fields. Here this conflict is removed by replacing global scalar fields with local scalar fields. These are represented by  $\bar{S}_{x}$ where $x$ is any location in Euclidean space or space time. Here  $S$ represents  the different type of numbers, (natural, integers, rational , real, and complex).

           The association of scalar fields with vector spaces and the Yang Mills observation raises the question, What corresponds to the Yang Mills observation for numbers? The answer is that two different concepts, number and number meaning or value, are conflated in the usual use of mathematics. These two concepts are distinct.

             \section{Background}

             \subsection{Mathematics}\label{M}
             Studying some consequences of the separation of number value from number requires a descriptipon of the elements of mathematics.  Here mathematics  is considered to include structures \cite{Shapiro} or models \cite{Keisler} of systems  of many different types and the relations between them.   A structure consists of a base set, a few basic operation, none or a few relations and constants. The structure must satisfy  a set of axioms appropriate for the structure  type. Expansion of the brief description given here on the distinction between number and number value and consequences for physics are given elsewhere  \cite{BenIJTP,BenINTECH}.

             A few of the many examples are the natural number structure, $\bar{N}=\{N, + \times, <,0,1\}$, The rational number structure, $\overline{Ra}=\{Ra,+,-,\times,\div,<,0,1\}$, the real number structure, $\bar{R}=\{R,+,-,\times,\div,<,0,1,\}$,  the complex number structure,, $\bar{C}=\{C,+,-,\times,\div,0,1\}$, normed vector spaces, denoted collectively by $\bar{V}=\{V,+,-,\odot, |-|,\vec{v}\}.$   Here $\odot$ and $\vec{v}$ denote scalar vector multiplication and an arbitrary vector. Symbols with an overline denote a structure, the same symbols without an overline denote base sets.

           \subsection{Expansion}
           The description of mathematics structures given above is a special case of a more general description.  For each type of number and vector space there are many different structures or models.  The structures differ by scaling factors.\footnote{These expansions are examples of nonDiophantine arithmetic \cite{Burgin}.} The description of the expansion for each number type will be brief. For details see the authors references listed.

           \subsubsection{Natural numbers}

           The natural number structure containing all natural numbers is denoted by $\bar{N}.$ The natural number structure consisting of all even numbers is $\bar{N}^{2}=\{N_{2},+_{2},\times_{2},<_{2},0_{2},1_{2}\}.$ The structure containing every $nth$ number is $\bar{N}^{n}=\{N_{n},+_{n},\times_{n},<_{n},0_{n},1_{n}\}.$  The usual structure $\bar{N}$ is identified with $\bar{N}^{1}.$ That is $0\equiv 0_{1},1\equiv 1_{1},2\equiv 2_{1},\cdots.$  The superscripts and subscripts $2$ and $n$ will be referred to as scale or valuation factors.

           In these structures, $N_{n}$ is the set of every $nth$ number in $N_{1}$.  Values of the numbers are determined by their position  in the well ordering. For example $0$ and $n$ are the  base set numbers that have values $0$ and $1$ in $\bar{N}^{n}.$   This well ordering can be expressed  by the use of a valuation function. This function associates values with numbers in different number structures.

           Let $v_{n}$ be the valuation function of numbers in $\bar{N}^{n}.$  The domain of $v_{m}$ is $N_{n}$, the set of every $nth$ number in $N=N^{1}.$ One has
           \begin{equation}\label{vn0}v_{n}(0)=0,\;v_{n}(n)=1,\cdots v_{n}(m)= \frac{m}{n},\cdots.\end{equation}For each number $m$ in $\bar{N}^{n}$ $v_{n}(m)$ is the value or meaning of $m$. Note that $v_{n}$ is defined only on numbers containing $n$ as a factor.

           These relations show that number value is different from number in all structures $\bar{N}^{n}$ except the structure where $n=1.$ In $\bar{N}^{1}=\bar{N}$ number and number value are identified.  This can be expressed by  \begin{equation}\label{v1mm}
           v_{1}(m)=m\end{equation}for all base set numbers, $m$.

           Another representation of numbers and their values is as $m_{n}$ This is the number with value $v_{n}=m$ in $\bar{N}^{n}.$  The only number with number and number value conflated for all valuation factors is $0$.  One has $v_{n}(0)=0$ for all $n$.

           The requirement that the structures, $\bar{N}^{n}$ satisfy the axioms of arithmetic gives relations between $+_{n}, \times_{n},<_{n},0_{n},1_{n}$ and  $+_{1}, \times_{1},<_{1},0_{1},1_{1}$  that must be satisfied.  These are $$+_{n}\equiv +_{1},\times_{n}\equiv\frac{\times_{1}}{n},<_{n}\equiv <_{1},0_{n}\equiv 0,1_{n}\equiv [n]_{1}=n.$$ The  components of the structure, $\bar{N}^{n}$ are related to the components of $\bar{N}^{m}$, by  $$+_{n}\equiv +_{m},\times_{n}\equiv[\frac{m}{n}\times_{m}]_{n},<_{n}\equiv <_{m},0_{n}\equiv [\frac{n}{m}0]_{m}=0,1_{n}\equiv [\frac{n}{m}1]_{m}.$$

           The nomenclature used here will  be  much used in this work.  For arithmetic combinations of numbers, square brackets are used to separate the value of the expression from the scale factor of the number structure containing the number.  For example $[a\times c]_{b}$ is the expression in the number structure scaled by  $b$ that has value $a\times c.$

           Expansion of integer number structures will not be discussed separately because the description is almost the same as that for natural numbers. Structures for integers differ from those for natural numbers in that they include the subtraction operation.

            \subsubsection{Rational  numbers}
            Let $\overline{Ra}^{s}=\{Ra,+_{s},-_{s},\times_{s},\div_{s},<_{s},0_{s},1_{s}\}$ represent the rational number structure scaled by the positive rational number, $s$. Unlike the case for the natural numbers and integers  the base set $Ra$ is the same for all scaled rational number structures.

            It follows from this that a rational number, as a base set element, has no intrinsic value. Its value is determined by the scaling or value factor of the associated structure. For example the rational number $27/32$ in $Ra$  has value $v_{s}(27/32)$ in $\bar{Ra}^{s}.$

            Extension of  Eq. \ref{v1mm} to rationl numbers, shows that  number and number value are identified in $\overline{Ra}^{1}$. Then  $v_{1}(27/32)=27/32.$ For every pair, $i,j$ of integers in the base set $Ra$ of $\bar{Ra}^{1}$\begin{equation}\label{v1ij}v_{1}(i/j) =i/j.\end{equation}

            In $\overline{Ra}^{s}$ the base set number, $i/j$ is represented by the number $[v_{s}(i/j)]_{s}.$ From \begin{equation}\label{vsij}v_{s}(i/j)=\frac{1} {s}v_{1}(\frac{i}{j}) =\frac{i}{sj} \end{equation} one has $[i/sj]_{s}$ is the same number in $\overline{Ra}^{s}$ as  $[i/j]_{1} $ is in $\overline{Ra}.$

            As is seen from the above, rational numbers in $\bar{Ra}^{s}$ can also be represented in the form $a_{s}.$  This is the number with value $a$ in $\bar{Ra}^{s}.$  The rational number $-4.32$ in $Ra$ has value $v_{s}(-4.32)$ in $\bar{Ra}^{s}.$ The representation in the form $a_{s}$ of this number is $[v_{s}(-4.32)]_{s}.$  If one assumes that number and number value are conflated in $\overline{Ra}^{1},$ as in Eq. \ref{vsij}, then $[v_{s}(-4.32)]_{s}=[-4.32/s]_{s}$.

            Let $\overline{Ra}^{t}$ be a rational number structure scaled by the positive rational factor, $t.$ The  operations,  order relation and constants in $\overline{Ra}^{t}$ can be mapped by a number preserving value changing map into the corresponding components of $\overline{Ra}^{s}$ as in \begin{eqnarray}\label{Rats}
            a_{t}\rightarrow [\frac{t}{s}a]_{s},\; \pm_{t}\rightarrow \pm_{s},\; \times_{t}\rightarrow [\frac{s}{t}\times]_{s},\; \div_{t}\rightarrow[\frac{t}{s}\div]_{s},\nonumber\\ <_{t}\rightarrow <_{s},\; 0_{t}\rightarrow 0_{s}=0,\; 1_{t}\rightarrow [\frac{t}{s}1]_{s}.\hspace{2cm} \end{eqnarray}

              \subsubsection{Real and complex numbers}

            The  relations between rational number structures also hold for real  and complex number structures. Let $\bar{R}^{s}$ and $\bar{R}^{t}$ be two real number structures and $\bar{C}^{s}$ and $\bar{C}^{t}$ betwo complex numkber structures. Here $s$ and $t$ are two positive real number scaling factors. The scale factors for complex numbers are restricted here to be positive real numbers.

            Let $S$ denote either the  rational, real, or complex numbers. Then $\bar{S}^{s}$ denotes a  rational, real, or complex number structure with $s$ a real scaling factor. The number preserving value changing map of $\bar{S}^{t}$ onto $\bar{S}^{s}$ is given by  Eq. \ref{Rats}. The order relation is missing for complex numbers. Division and subtraction is missing from natural numbers.  The scaling used in this work is a special case of that described by Czachor \cite{Czachor}.

             The number preserving value changing map can be represented by a  connection that maps components of $\bar{S}^{t}$ into  $\bar{S}^{t}_{s}.$  This is shown explicitly by \begin{equation}\label{Cst}C(s,t)\bar{S}^{t}=\bar{S}^{t}_{s}
            \end{equation}where \begin{equation}\label{CSts}\bar{S}^{t}_{s}=\{S,\pm_{s}, [\frac{s}{t}\times]_{s}, [\frac{t}{s}\div]_{s},<_{s},0_{s},[\frac{t}{s}1]_{s}\}. \end{equation} The order relation is missing from components of the complex number structure.  Here $\bar{S}^{t}_{s}$ is the number structure whose components represent the components of $\bar{S}^{t}$ in terms of those of $\bar{S}^{s}.$
            The structure, $\bar{S}^{t}_{s}$ satisfies the same axioms as do $\bar{S}^{t}$ and $\bar{S}^{s}.$

            The number $0$ is the only number whose value is unaffected by the connection.  The value $0$ of the number $0_{s}$ for any $s$ is $0$. It is the only number that can be conflated with its value, $0_{s}\equiv 0$ for all $s.$ In a sense it is the "number vacuum".

            A caveat worth noting is that the connection does not commute with multiplication or division. To see this one has \begin{equation}\label{Cstatb}C(s,t)[a\times b]_{t}= C(s,t)(a_{t})C(s,t)(\times_{t})C(s,t)(b_{t})=[\frac{t}{s}]_{s}[a\times b]_{s}.\end{equation}One of the two $t/s$ factors coming from $C(s,t)(a_{t})$ and $C(s,t)(b_{t})$ is canceled by the factor $s/t$ coming from $C(s,t)(\times_{t}).$ Here multiplication acts before connection.

            Reversing the order of operations gives \begin{equation}\label{Cstta}C(s,t)(a_{t})\times_{s}C(s,t)(b_{t})= [(\frac{t}{s})^{2}]_{s}[a\times b]_{s}.\end{equation} Comparison of this result with $C(s,t)[a\times b]_{t}$ shows the scale factor $(t/s)^{2}$ instead of $t/s.$ Replacing the multiplication operation with division gives the scale factor  $1$ instead of $t/s.$

            There is a distinction to be made between numbers as meaningless symbol strings and numbers having meanings in expressions such as those used in physics.  For example the number $4.56$ as  a rational  number  symbol string  has value $(1/b)4.56$ in the structure $\bar{S}^{b}.$  Note that for $b=1$ number and number value coincide.  The connection $C(d,b)$ maps $(1/b)4.56$ to the number \begin{equation}\label{Cb4}C(d ,b)(1/b)4.56 =(b/d)(1/b)4.56=(1/d)4.56. \end{equation}in the structure $\bar{S}^{d}.$

            In expressions in which numbers have meaning, they are   expressed in the form, $a_{b}$  as a number with value or meaning\footnote{Meaning and value of numbers, vectors, and numerical or vectorial physical quantities are equivalent.} $a$ in $\bar{S}^{b}.$  For example  $4.56_{b}$ is the rational number with value $4.56$ in $\bar{S}^{b}.$ For numbers in $\bar{S}^{d}$ whose meaning expressions involve  few symbols, square brackets are used  to separate meaning from scaling factors. Example:  the number whose meaning is expressed by $x\times y$ in $\bar{S}^{d}$ is written as $[x\times y]_{d}.$

            Use of the connection to map the number $a_{b}$ to a number in $\bar{S}^{d}$ gives \begin{equation}\label{Cd4}C(d,b) a_{b}=[(b/d)a]_{d}.\end{equation} This differs from Eq. \ref{Cb4} by the presence of the factor $b$. Since all numbers appearing in physical expressions have meaning,  numbers appearing in these expressions have the form of $a_{b}.$

            \subsubsection{Vector spaces}

            As was the case for scalars there are scaled vector spaces, $\bar{V}^{s}$ where $$\bar{V}^{s}=\{V_{s},\pm_{s},\odot_{s}, |-|_{s},\vec{v}_{s}\}.$$ The scalar vector product, $\odot_{s}$, is a map from numbers in $\bar{S}^{s}$ and vectors in $\bar{V}^{s}$ to vectors in $\bar{V}^{s}.$ The norm maps vectors  in $\bar{V}^{s}$ to numbers in $\bar{S}^{s}.$ The scale factor for vector spaces is the same as that for the associated scalars.

            The relation between $\bar{V}^{s}$ and $$\bar{V}^{t}=\{V_{t},\pm_{t},\odot_{t}, |-|_{t},\vec{v}_{t}\}$$ is similar to that between the number structures, $\bar{S}^{s}$ and $\bar{S}^{t}.$ The connection $C(s,t)$, maps $\bar{V}^{t}$ to $\bar{V}^{t}_{s}$, a representation of the components of $\bar{V}^{t}$ in terms of those of $\bar{V}^{s}.$  The components of $\bar{V}^{t}_{s}$ are shown in \begin{equation}\label{Vts} \bar{V}^{t}_{s} =\{V_{s},\pm_{s},[\frac{s}{t}\odot]_{s} ,[\frac{t}{s}|-|]_{s}, [\frac{t}{s}\vec{v}]_{s}\}.\end{equation}

           \section{Local mathematics}

           Mathematics is usually considered to be global. Structures  for different types of mathematical systems are not assigned locations in space time.  They are present  outside of space, time, or space time.

           In this work global mathematics is replaced by  local mathematics. Local mathematics consists of structures of different types of mathematical systems at each point in  space time.

           Let $\mathbb{S}$ denote space time.  For each location, $x$ in $\mathbb{S},$ mathematical  structures at $x$ have  $x$ as a location subscript.  Examples include  $\bar{N}_{x},\bar{C}_{x}, \bar{V}_{x}$ Other types of mathematical systems that use scalars in their axiomatic description are also included.

           Define $M_{x}$ to be the collection at $x$ of structures of all types of mathematical systems that include scalars in their descriptions.  The structure collection includes numbers of different types, vector spaces, operator algebras, group representations, and many other system types.

           The mathematics of local structures is completely independent of their location. The mathematics in  $M_{y}$ at $y$ is the same as the mathematics in $M_{x}$ at $x$. Locations of structures have no effect on their meaning or use in mathematics. Numbers, vectors and their values and arithmetic combinations of numbers and vectors can be identified with their value or meaning.

           This is shown by noting that $M_{x}=M_{x}^{1}$ The mathematical structures in $M^{1}_{x}$ include $\bar{N}^{1}, \overline{Ra}^{1}_{x},\bar{R}^{1}_{x}, \bar{C}^{1}_{x},\bar{V}^{1}_{x}$ and other types of structures that include numbers in their description. Numbers, vectors and arithmetic combinations can be identified with their values or meanings.  For example the rational number $4.64_{1}\equiv 4.64$, the arithmetic combination, $[aOb]_{1}\equiv aOb, \vec{v}_{1}\equiv\vec{v}.$ Here $O$ denotes any of the four operations, $\pm,\times, \div.$

            These considerations show that the local mathematics at all locations is equivalent to global mathematics. This  can be succinctly represented by \begin{equation}\label{M1x}M= \bigcup_{x\epsilon \mathbb{S}} M^{1}_{x}.\end{equation}Here $M$ without superscript or subscript denotes global mathematics.

           \subsection{Meaning or value fields}

           The introduction  of a  space time dependent meaning or value field,\footnote{Meaning field and value field are different names for the same field. The names express the fact that numbers have value and meaning.} $g$, on $\mathbb{S}$ generalizes the description of local mathematics. In general Eq. \ref{M1x} is not valid.  Meanings or values of mathematical elements in different types of structures at different locations depend on the values of $g$ at the structure locations.  The structure collection $M_{x}$ at $x$ becomes the collection $M^{g(x)}_{x}$.  Components include scalar structures of different types, represented generically by $\bar{S}^{g(x)}_{x},$ vector spaces, $\bar{V}^{g(x)}_{x},$ and structures for other types of systems that include scalars in their description.

           The relation between global and local mathematics depends on the $g$ field.  In the special case that $g(x)=1$ everywhere, then Eq, \ref{M1x} is valid.  If $g(x)=1$ in a region $\mathbb{R}$ of $\mathbb{S}$, then  mathematics is global in $\mathbb{R}$,
           \begin{equation}\label{MbR}M^{\mathbb{R}}=\bigcup_{x\epsilon \mathbb{R}}M^{1}_{x}.
           \end{equation} In general if $g(x)$is a positive constant for all $x$ in $\mathbb{R},$ then mathematics is global in $\mathbb{R}.$

           The presence of the $g$ field  has the result that the values or meanings of numbers, vectors, and their combinations depend on their location in space time. This dependence can be seen explicitly by generalizing the definition of the connection used in Eq. \ref{Cst}.

           \subsection{Connections in space time}

           The action of $C_{g}(x,y)$  as  a number preservinng value changing map on numbers in $\bar{S}^{g(y)}_{y}$ and vectors in $\bar{V}^{g(y)}_{y}$ is shown by \begin{equation}\label{Cxyag}
           C_{g}(x,y)a_{g(y)}=[\frac{g(y)}{g(x)}]_{g(x)}a_{g(x)}=[\frac{g(y)}{g(x)}a]_{g(x)}
           \end{equation}and\begin{equation}\label{Cxyvg}C_{g}(x,y)\vec{v}_{g(y)}=[\frac{g(y)} {g(x)}]_{g(x)}\vec{v}_{g(x)}=[\frac{g(y)}{g(x)}\vec{v}]_{g(x)}.\end{equation} For
           arithmetic combinations of numbers \begin{eqnarray}\label{CxyaOb}C_{g}(x,y) [aOb]_{g(y)}=C_{g}(x,y) a_{g(y)}C_{g}(x,y)O_{g(y)}C_{g}(x,y)b_{g(y)}\nonumber\\ =[\frac{g(y)} {g(x)}(aOb)]_{g(x)}.\hspace{4cm}\end{eqnarray}If $O=\pm$, $C_{g}(x,y)\pm_{g(y)}=\pm_{g(x)}.$ If $O=\times$, $C_{g}(x,y)\times_{g(y)}=[\frac{g(x)} {g(y)}\times]_{g(x)}.$ If $O=\div$, $C_{g}(x,y)\div_{g(y)}=[\frac{g(y)}{g(x)} \div]_{g(x)}.$

            For scalar vector multiplication one has\begin{eqnarray}\label{CxyaOv}C_{g} (x,y)[a\odot \vec{v}]_{g(y)} =C_{g}(x,y)\vec{v}_{g(y)}C_{g}(x,y)\odot_{g(y)} C_{g}(x,y)\vec{v}_{g(y)} \nonumber\\ =[\frac{g(y)}{g(x)}a\odot\vec{v}]_{g(x)}. \hspace{4cm}\end{eqnarray}

            These maps can be collected together by representing $C_{g}(x,y)$ on mathematical structures. For number structures and vector spaces $C_{g}(x,y) \bar{S}^{g(y)}_{y}= \bar{S}^{g(y)} _{x}$ and $C_{g}(x,y) \bar{V}^{g(y)}_{y}= \bar{V}^{g(y)} _{x}.$  The structures $\bar{S}^{g(y)}_{x}$ and $\bar{V}^{g(t)}_{x}$ with their components are shown in Eqs. \ref{CSts}  and \ref{Vts} where $t=g(y)$ and $s=g(x).$

           The action of $C_{g}(x,y)$ on scalars, vectors and their  combinations will be referred to as a parallel transport of these elements from one location to another. It is   similar to the action of unitary gauge transformations as parallel transports of vectors  in gauge theories \cite{Montvay,Mack}.

           As Eq. \ref{Vts} shows parallel transport of a vector multiplies it by a number.  This suggests a similarity to conformal transformations \cite{Ginsparg}.  Parallel transport is different in that, unlike conformal transformations, numbers and arithmetic combinations of numbers are also scaled.  Scalar products of vectors and trigonometric functions are scaled just like numbers. $$C_{g}(x,y) [\vec{u}\cdot\vec{v}]_{g(y)} =[\frac{g(y)}{g(x)} \vec{u}\cdot\vec{v}]_{g(x)}$$  and as a trigonometric example $$C_{g}(x,y)[\cos(\theta)]_{g(y)}=[\frac{g(y)} {g(x)}\cos(\theta)]_{g(x)}.$$

           \subsection{Local Mathematics with the $g$ field}

           With the $g$ field present the definition of local mathematics  is a generalization of that of $M^{1}$ in Eq. \ref{M1x}.  Here $M^{g}$ denotes the local mathematics at all points of space time where \begin{equation}\label{Mqx}M^{g}= \bigcup_{x\epsilon \mathbb{S}} M^{g(x)}_{x}.\end{equation}

           Here and from now on the  exponential form of the $g$ field as in \begin{equation}\label{gal}g(x)=e^{\alpha(x)}\end{equation} will be used. In this case $\alpha$ will be referred to as the meaning  or value field for numbers, vectors, and other types of mathematical elements. The ratios $g(y)/g(x)$ in the above equations become $e^{-\alpha(x)+\alpha(y)}.$  The exception to this change of representation is the use of $g(x)$ as a subscript or a superscript for mathematical elements and structures of different types.  This done to make notation less clumsy.

           \subsection{The scalar field, $\alpha$}

           The scalar field $\alpha$ is basic to all that follows. The exponential of $\alpha$  determines the meaning or value of scalar properties of mathematical  elements  and physical systems. It determines the value  of numbers and scalar properties of vectors  (such as the length) at different locations in space time. The meaning or value of a number at  $y$  is different from the value or meaning of the same number at $x$  If $r_{g(y)}$ is a number with value $r$ at location $y$, the value of this number at $x$ is $[e^{-\alpha(x)+\alpha(y)}r]_{g(x)}$.

          The transport factor $e^{-\alpha(x)+\alpha(y)}$ is the scalar equivalent of the unitary gauge transformation, $U(x,y),$ from $y$ to $x$ for vectors  in gauge theories. Just as $U(x,y)\psi(y) $ is the same vector in $\bar{V}^{g(x)}$ as $\psi(y)$ is in $\bar{V}^{g(y)},$  the scalar, $[e^{-\alpha(x)+\alpha(y)}r]_{g(x)}$ is the same number in $\bar{S}^{g(x)}$ as $r_{g(y)}$ is in $\bar{S}^{g(y)}.$  Just as  $U(x,y)\psi(y)$ is different from $\psi(x)$, so is the value, $e^{-\alpha(x)+\alpha(y)}r$ different from $r.$

           \section{Effect of local mathematics and $\alpha$ in physics and geometry}

           Local mathematics, the meaning or value field, and the connection or parallel transport operator provide the arena or background  for theoretical descriptions of properties of  physical systems, computer outputs and experimental outcomes. These   show a  dependence on variations in the $\alpha$ field. This is a consequence of the fact that they are meaningful and have value.

           This emphasis on the meanings or values of numerical or vectorial physical quantities has the consequence that the effect of local mathematics and $\alpha$ show up in any quantity expressed as a scalar, vector, etc. valued function over space, time, or space  time. They also show up in any theory prediction or a computer output and any experimental outcome. The effect shows up in a different form in computer outputs and experimental outcomes than in theory expressions.

           \subsection{\label{ECET}Effect of  $\alpha$ on computers, experiment, and theory}

           As shown in Section \ref{M} numbers and vectors can be expressed in two forms.  A base set number $r$ at location $y$ in $\bar{S}^{g(y)}_{y}$ can be expressed as $r/e^{g(y)}$ or as $a/ g(y).$ Here $a$ is the value of $r$ in $\bar{S}^{g(y)}_{y}$.

           These two forms affect  computer outputs, experimental outcomes and theory descriptions differently.  Outputs of computers and experimental outcomes are numbers produced by thesse operations. A number, $r$ created at location $y$, as a symbol string representing a number in $\bar{S}^{g(y)}$, has value $r/e^{\alpha(y)}$.

           All theory expressions have meaning.  Numbers, number variables, scalar fields, and theory formulas and equations all  have meaning.  For this reason numbers, number variables and any representation of nummbers in theory must have the form, $a_{b}.$ Theory formula showing numbers, number, variables and representations of numbers at some space time location $y=\textbf{y},s$ are represented by $a_{g(y)}$.  Scalar functions or fields, classical or quantum, $f(y),$ become $f(y)_{g(y)}.$  The location variable is in both the $g$ and $f$ functions.

           This paper is limited to physical and geometric properties  represented by functions of one variable. Properties of physical systems described by functions whose domain is pairs, triples, or $n$-tuples of space, time, or space time locations are not included. A simple example  is the evolution of a wave function describing the interaction between two particles. Entangled states of particles are another example.\footnote{One way to treat these properties is described in \cite{BenENT} for entangled states.}

           \subsection{effect of $\alpha$ on space time integrals and derivatives}

           The use of local mathematics and the value field, $\alpha,$ affects theoretical deescriptions of many physical properties. This is especially so for properties that are represented by integrals or derivatives over space, time, or space time.

            Let $f$ be a scalar or vector field on space time.  For each $y$ $f(y)_{g(y)}$   is a scalar in $\bar{S}^{g(y)}_{y}$ or\footnote{Recall that $\bar{S}$ denotes a generic structure for different number types, usually real or complex numbers.} a vector in $\bar{V}^{g(y)}_{y}.$ The integral $\int f(y)_{g(y)}dy$ is not defined.  The definition of the integral is the limit of sums of the integrand $f(y)_{g(y)}$ over values of $y$.  The implied addition of numbers or vectors, in different structures at different locations, is not defined. Arithmetic operations are defined only within a structure, not between structures.  An equivalent condition is that the argument of the $g$ function must not be an integration variable.

            This is fixed by parallel transport of the integrands to a reference location, $x$, before integration. Use of Eq. \ref{CxyaOb} and a similar equation for vectors gives  \begin{equation}\label{Ifgx}I(f)_{g(x)}=[e^{-\alpha(x)}\int e^{\alpha(y)} f(y)dy]_{g(x)}.\end{equation}The price of the use of local mathematics and $\alpha$ is the presence of the exponential factors multiplying the integrand.

            The same problem  that causes integrals to be  undefined also holds for derivatives.  The definition of derivatives involves comparison of values of  $f$, $f(y+d^{\mu}y)_{g(y+d^{\mu}y)}$ and $f(y)_{g(y)},$ at neighboring space time points. These comparisons are not defined. The problem is fixed by parallel transport of $f(y+d^{\mu}y)_{g(y+d^{\mu}y)}$ to $f(y+d^{\mu}y)_{g(y)}$.

            The $\mu$ component of the resulting derivative of $f$ at $y$  is $D_{\mu,y}f$ where\begin{eqnarray} \label{Dmeta} D_{\mu,y} f(y)_{g(y)} =[\frac{e^{-\alpha(y) +\alpha (y+d_{\mu}y)} f(y+d_{\mu}y) -f(y)} {d_{\mu}y}]_{g(y)}\nonumber\\=[ (\partial_{\mu,y}+ A_{\mu}(y)) f(y)]_{g(y)}.\hspace{1cm} \end{eqnarray}The exponential factor comes from parallel transport of $[f(y+d^{\mu}y)]_{g(y+d^{\mu}y)}$ at $y+d_{\mu}y$ to $[f(y+d_{\mu}y) ]_{g(y)}$ at $y$. The limit, $\lim_{d_{\mu}y\rightarrow 0}$ is understood.

            The second line of the equation is obtained by Taylor expansion of $\alpha(y+d_{\mu}y)$  and expansion of the exponential to first order in small quantities. The gradient vector field $\vec{A}(y)$ has four components \begin{equation}\label{umAy}A_{\mu}(y)= \frac{d_{\mu}\alpha(y)}{dy}. \end{equation} Note that if $f(y)_{g(y)}=[e^{-\alpha(y)}]_{g(y)}$ is the number with the same value as the scaling factor, then\footnote{see Eq. \ref{Dmeta}}
            Eqs. \ref{CxyaOb} and \ref{Dmeta} are defined  for scalar fields.  They also apply to vector fields.  An example will be given later on.

        The derivations of the expressions for integrals and derivative are based on $\vec{A}$ as a gradient vector field. A more general approach starts with a vector field, $\vec{A}.$ The above equations with $e^{-\alpha(x)+\alpha(y)}$ as the connection factor are valid if $\vec{A}$ is integrable.  If $\vec{A}$ is not integrable, the  component connections must be replaced by path dependent factors.\footnote{The Bohm Aharonov effect \cite{AhBo} is a quantum mechanical example of the effect of nonintegrable vector fields.}

         Differential geometries based on a nonintegrable vector field  were described more than $100$ years ago by Weyl \cite{Weyl}. Weyl introduced  a nonintegrable real vector field to describe the scaling  under parallel transfer of quantities.\footnote{A good description of the work, including  criticism by Einstein  and subsequent developments of early gauge theory are in a book by O'Raifeartaigh \cite{OR}.} The $\alpha$ dependent scaling  used here is quite different from that of Weyl as it is based  on local mathematics.   Mathematical structures are also scaled by a space and time dependent scale factor.

         Use of nonintegrable vector fields to describe parallel transports of numerical and vector quantities in a geometry based on local mathematics would be complicated.  The geometry would have to be able to include  path dependent parallel transports. Whether this is possible or not must await further work.

           \section{The geometry, $\mathbb{G}^{\alpha}$}

           \subsection{Introduction}

           Variations in the value field affect both physics and the geometry in which physical systems move.  Perhaps the most direct way to see this is to show the effect of the $\alpha$ field on metric tensors.  The relation between geometries with and without the presence of  the $\alpha$ field is  shown by the relation between the metric tensors for the geometries.

           Let $\mathbb{G}$ be the geometry based on the metric tensor $\gamma_{\mu,\nu}(y)$.  The corresponding geometry $\mathbb{G}^{\alpha}$ that includes the effect of $\alpha$ begins with the tensor  $[\gamma_{\mu,\nu}(y)]_{g(y)}.$ For each $y$ the tensor is an element in the mathematics, $M^{g(y)}$ at $y.$. Use of this tensor in  theory expressions to describe properties of a geometry is not defined because arithmetic combinations or comparisons are in different mathematicl structures.  This is fixed by parallel transport to a  mathematics, $M^{g(x)},$ at a reference location, $x.$

           The metric tensor for the geometry $\mathbb{G}^{\alpha}$ is is given by $[e^{-\alpha(x)+\alpha(y)}\gamma_{\mu,\nu}(y)]_{g(x)}.$ This geometry differs from $\mathbb{G}$ by the presence of the $y$ dependent exponential factor, $e^{\alpha(y)}$ and the reference location factor, $e^{-\alpha(x)}.$

           A simple example is obtained by setting \begin{equation} \label{gest}[\gamma( y)]_{g(y)}= \eta(y)]_{g(y)}=[-1,1,1,1]_{g(y)}\end{equation} where $\eta(y)_{g(y)}$ is the metric tensor for space time.  Parallel transport of the metric tensor  to a single reference location, $x$  gives \begin{equation}\label{eaxeta}
           [e^{-\alpha(x)+\alpha(y)}\eta(y)]_{g(x)}=[e^{-\alpha(x)+\alpha(y)}(-1,1,1,1)]_{g(y)}.
           \end{equation} This tensor is still diagonal in the indicies.  The only difference between it and $\eta(y)$ is multiplication by a location dependent scalar, $e^{-\alpha(x)+\alpha(y)}.$

          The geometry, $\mathbb{G}^{\alpha}$, represented by  the metric tensor of Eq. \ref{eaxeta}, is quite flexible. It is a flat space time if and only if $\alpha(y)$ is constant everywhere.\footnote{This description of $\alpha$ dependent geometry also applies to $3$ dimensional Euclidean space with $\delta_{j,k}$ as the metric tensor.}  The geometry is not flat if $\alpha(y)$ is variable. The rate of deviation  from flatness is determined by the vector gradient field, $\vec{A}$, Eq. \ref{umAy}.

          One sees from this that variations in $\alpha$ determine the deviation of $\mathbb{G}^{\alpha}$ from flatness.  The gradient, $\vec{A},$ of $\alpha$ determines the  amount of deviation. If $\vec{A}=0$ everywhere then $\mathbb{G}^{\alpha}$ is the usual flat space time.  If $\vec{A}(y)$  is large  for all $y$ in some region, then $\mathbb{G}^{\alpha}$ is quite distorted in the region.
          \subsection{Geodesics}

           The path followed by a free particle in $\mathbb{G}^{\alpha}$ is a geodesic. This path is determined by the geodesic equation. The path  maximizes \cite{Carroll} the proper time for a particle moving on the path.

           The form of the geodesic equation in curved space time is  given by   \begin{equation}\label{gdscG}\frac{d^{2}p^{\mu}}{d\tau^{2}}+ \Gamma^{\mu}_{\nu,\rho} \frac{dp^{\nu}}{d\tau}\frac{dp^{\rho}}{d\tau}=0 \end{equation}where \cite{Weinberg} \begin{equation}\label{Gmnr}\Gamma^{\mu}_{\nu,\rho}=\frac{1}{2}g^{\mu\sigma} (\partial_{\nu}g_{\rho\sigma}+\partial_{\rho}g_{\sigma\nu}-\partial_{\sigma}g_{\nu\rho}). \end{equation} The metric tensor is denoted by $g_{\mu,\nu}.$

           If $g_{\mu,\nu}$ is diagonal in the indices then $\Gamma^{\mu}_{\nu,\rho}$ becomes $$\frac{1}{2}g^{\mu,\mu}\delta_{\mu,\sigma}(\partial_{\nu}g_{\mu,\mu}\delta_{\rho,\sigma} +\partial_{\rho} g_{\mu,\mu}\delta_{\nu,\mu} -\partial _{\mu}g_{\nu,\nu}\delta_{\rho,\nu})$$  The geodesic equation becomes  \begin{eqnarray}\label{gdsG}\frac{d^{2}p^{\mu}}{d\tau^{2}}+  \frac{1}{2}g^{\mu,\mu}\delta_{\mu,\sigma}(\partial_{\nu}g_{\mu,\mu}\delta_{\rho,\sigma} \hspace{2cm}\nonumber\\+\partial_{\rho} g_{\mu,\mu}\delta_{\nu,\mu} -\partial _{\mu}g_{\nu,\nu}\delta_{\rho,\nu}) \frac{dp^{\nu}}{d\tau}\frac{dp^{\rho}}{d\tau}=0 \end{eqnarray}From this one obtains  \begin{eqnarray}\label{gdsa} \frac{d^{2}p^{\mu}}{d\tau^{2}}+  \frac{1}{2}g^{\mu,\mu}(\partial_{\nu}g_{\mu,\mu} \frac{dp^{\nu}}{d\tau}\frac{dp^{\mu}}{d\tau} \hspace{2cm}\nonumber\\+\partial_{\rho} g_{\mu,\mu}\frac{dp^{\mu}}{d\tau}\frac{dp^{\rho}}{d\tau} -\partial _{\mu}g_{\nu,\nu} \frac{dp^{\nu}}{d\tau}\frac{dp^{\nu}}{d\tau})=0 \end{eqnarray}

           Use of   \begin{equation}\label{gtmumu}g_{\mu,\mu}=\gamma_{\mu,\mu} (p(\tau)) = e^{-\alpha(x)+\alpha(p(\tau))} \eta_{\mu,\mu},\end{equation}and \begin{equation}\label{gbmumu}g^{\mu,\mu}=\gamma^{\mu,\mu}(p(\tau))=e^{\alpha(x) -\alpha(p(\tau))}\eta^{\mu,\mu}\end{equation} gives\begin{eqnarray}\label{Aeta} \frac{d^{2}p^{\mu}}{d\tau^{2}}+\frac{1}{2}(A_{\nu}(p(\tau)) \frac{dp^{\nu}}{d\tau}\frac{dp^{\mu}}{d\tau}+A_{\rho}(p(\tau))\frac{dp^{\mu}}{d\tau} \frac{dp^{\rho}}{d\tau}\nonumber\\-\eta^{\mu,\mu}\eta_{\nu,\nu}A_{\mu}(p(\tau)) \frac{dp^{\nu}}{d\tau}\frac{dp^{\nu}}{d\tau})=0.\hspace{1cm}\end{eqnarray}Replacing $\rho$ with $\nu$ gives \begin{eqnarray}\label{etmunu}\frac{d^{2}p^{\mu}}{d\tau^{2}} = -A_{\nu}(p(\tau)) \frac{dp^{\nu}}{d\tau}\frac{dp^{\mu}}{d\tau}+\frac{1}{2} \eta^{\mu,\mu} A_{\mu}(p(\tau))\eta_{\nu,\nu}\frac{dp^{\nu}}{d\tau} \frac{dp^{\nu}}{d\tau}\nonumber\\ = -A_{\nu}(p(\tau)) \frac{dp^{\nu}}{d\tau}\frac{dp^{\mu}}{d\tau}+\frac{1}{2} \eta^{\mu,\mu} A_{\mu}(p(\tau))c^{2}.\hspace{1.5cm}\end{eqnarray}

           The path taken by light or by  free falling particles is described by this equation.  It shows that the extent of the deviation of the path from a straight line is described by the components of the gradient vector field, $\vec{A}.$  If $\alpha$ is constant everywhere, the path is a straight line.

            Eq. \ref{etmunu} consists of number, vector and function values.  The corresponding expression for numbers, vectors and functions at any location, $x,$ is obtained by adding the  subscript $g (x)$ to each term. Since equations are invariant under parallel transport, change of any reference location $x$ to $x'$ has no effect on Eq. \ref{etmunu}.

            \subsection{The domain of $\mathbb{G}^{\alpha}$}

           The geometry $\mathbb{G}^{\alpha}$ can be used to describe a model cosmological universe. The domain extends from the time, $0$ of the big bang to the present at about $14$ billion years.  The spatial component of the domain consists of locations in the observable universe .

           The local physics done by us as observers takes place in a very small region, $\mathbb{R}$ of the universe.  If $x=t,\mathbf{x}$ where $t\approx 14$billion years and $\mathbf{x}$ is our spatial location in the universe, then $x$ can represent the location of $\mathbb{R}.$   Locations $w=s,\mathbf{w}$ in a local coordinate system with origin at $x$ are related to locations, $z$ in the universe by  $z=w+x.$

          The mathematical background  for the geometry, $\mathbb{G}^{\alpha}$ can be represented by local mathematical structures at each location in the  universe.  For each point $y=s,\mathbf{y}$ in the universe there is an associated local mathematics $M^{g(y)}.$

           This representation with  separate mathematical structures at each space time location in the background, is quite cumbersome. Another representation considers $\alpha$ to be a scalar field with space  components in  the flat Euclidean background\footnote{To keep things simple spherical and hyperbolic backgrounds are not described.} of the universe. For each location, $y$, $\alpha(y)$  determines the relation between number and number value, vector and vector value, etc.. The explicit relation for numbers and vectors is as follows:  If $r$  and $\vec{v}$ are a number and a vector at $y,$ their corresponding value or meaning is $e^{-\alpha(y)}r$ and $e^{-\alpha(y)}\vec{v}.$  If $\alpha(y)=0$, then $r$ and $\vec{v}$ represent a number and vector, and their values.  Number and vector are conflated with number value and vector value.

           The $\alpha$ determined relation between numbers and vectors and their corresponding values applies to physical quantities.  If the number $r$ represents the energy of a particle at $y$, the corresponding energy value is $e^{-\alpha(y)}r.$  If the vector $\vec{v}$ represents the momentum of a particle at $y$, then $e^{-\alpha(y)}\vec{v}$ represents the momentum value of the particle.

           The geometry $\mathbb{G}^{\alpha}$ is quite flexible in that each value field $\alpha$ describes a different model universe. The special case where $\vec{A}=0$  at all space and time locations corresponds to a flat space time and  global mathematics.  If $\alpha(y)=0$ everywhere also, number and vector coincide everywhere with number value and vector value.

            \section{Local physics in $\mathbb{G}^{\alpha}$}\label{LPG}

         As might be expected, variations in  $\alpha$ affect theory descriptions in local physics.  Local physics is the physics done by us at our location $x$ in $\mathbb{R}.$   The physics includes theory descriptions of the properties of local physical systems and theory support or refutation by experiments and measurements. Theory support or refutation takes place in $\mathbb{R}.$

         The effect of variations on $\alpha$ on theory descriptions of  properties of physical systems in $\mathbb{R}$ shows up in expressions that are integrals or derivatives of space, time, or spce time.  A summary of the effect is followed by some physical examples.

         \subsection{Classical particle dynamics}

         The motion of a free classical particle is determined by the geodesic equation, repeated here as \begin{equation}\label{gdeq}\frac{d^{2}p^{\mu}}{d\tau^{2}}= -A_{\nu}(p(\tau)) \frac{dp^{\nu}}{d\tau}\frac{dp^{\mu}}{d\tau}+\frac{1}{2} \eta^{\mu,\mu} A_{\mu}(p(\tau))c^{2}.\end{equation} The particle energy as a function of coordinate time, $s$ is given by the above equation with the proper time replaced by $s.$

         Use of the relation \begin{equation}\label{dtua}d\tau =\frac{ds}{\gamma(p(s))}
         \end{equation}and replacing $A_{\mu}(p(\tau))$  with $A_{\mu}(p(s))$ in Eq. \ref{gdeq} gives\begin{equation}\label{gsd}\frac{d}{ds}(\gamma(p(s))\frac{dp^{\mu}}{ds})= -A_{\nu}(p(s))\gamma(p(s)) \frac{dp^{\nu}}{ds}\frac{dp^{\mu}}{ds}+\frac{1}{2} \eta^{\mu,\mu} \frac{A_{\mu}(p(s))c^{2}}{\gamma(p(s))}.\end{equation}

         The relation between  the total particle energy and $\gamma$ is\begin{equation} \label{Ecm}E(p(s))=\gamma(p(s)) mc^{2}\end{equation} where $m$ is the rest mass of the particle.  Use of this relation in  Eq. \ref{gsd} with $\mu=0$ and $dp^{0}(s)/ds=c$  gives  \begin{equation} \label{Emps}\frac{1}{mc}\frac{d}{ds} E(p(s)) =\frac{1}{2}A_{0}(p(s)) \frac{mc^{4}}{E(p(s))}- A_{\mu}(p(s))\frac{dp^{\mu}(s)} {ds}\frac{E(p(s))}{mc}. \end{equation}

            The equation shows that the time dependent change of energy for a free particle  is independent of the particle mass.  From this one concludes that the time dependent energy  change is due to the  distortion of the underlying space time.   It is not due to the presence of any potential field or interaction with other particles.

             It should be noted that the terms in the above equations are values of numbers in $\bar{R}^{g(p(s))}_{p(s)}.$  The equations, expressed in numbers, have the subscript, $g(p(s))$ attached to each term.

             \subsection{Quantum mechanics}

             Quantum mechanics provides many examples of descriptions expressed by space and time integrals and derivatives \cite{BenENSQM}. Two examples are given here: the position expectation value of the wave function of a particle and the time dependent Schr\"{o}dinger equation.  The first example is an integral over space; the second is a time derivative.

          \subsubsection{Position expectation values}

           Let $\psi(t)$ be the nonrelativistic quantum state of a particle at time $t.$  The coordinate space components of this state are $[\psi(t,y)|y\rangle]_{g(y,t)}$.  Here $\psi(t,y)$ is the value of the amplitude, $\psi(t,y)_{g(y,t)}$. for finding  the system at $y$ in $\mathbb{G}^{\alpha}$ at time $t.$  The corresponding probability density is  $|\psi(t,y)|^{2}_{g(y,t)}.$

          The  position expectation at time $t,$  $\int [y|\psi(t,y)|^{2}]_{g(y,t)} dy_{g(y,t)},$  is not defined.  The integration variable is also the argument of $g$.   Parallel transport of the integrands to a common reference location, $x$ fixes the problem.  One obtains,$$[e^{-\alpha(x,t)} \int e^{\alpha(y,t)}y|\psi(y,t) |^{2}dy]_{g(x,t)}$$ for the position expectation value.

          This expression differs from the usual one by the presence of the $e^{\alpha(y,t)}$ factor in the integrand and  the presence of the factor, $e^{-\alpha(x,t)}$  multiplying the integral for the reference location, $x$.

          \subsubsection{Particle state dynamics in quantum mechanics}

          Another example of the effect of variations in $\alpha$ on properties of states of particles and their dynamics is given by the time dependent Schr\"{o}dinger equation. For this example the $\alpha$ field is restricted to depend on time only. For all spatial locations, $y$, $\alpha(y,t)=\alpha(t).$

          The Schr\"{o}dinger equation for a particle state $\psi(t)$ is \begin{equation} \label{Sch}[i\hbar]_{g(t)} \frac{d}{dt}[\psi(t)]_{g(t)}=[\tilde{H}]_{g(t)}[\psi (t)]_{g(t)}\end{equation}  where $\tilde{H}$ is the value of the Hamiltonian, $\tilde{H}_{g(t)}.$ A time varying $\alpha$ field and local mathematics has the result that the time derivative is not defined.  This is remedied by replacing $d/dt\psi(t)$ by $D_{t}\psi(t)$  where  by Eq. \ref{Dmeta} \begin{equation}\label{Dspt}[D_{t} \psi(t)]_{g(t)}=[(\frac{d}{dt}+A(t))\psi(t)]_{g(t)}.\end{equation} Here $A(t)=d\alpha(t) /dt.$  The Schr\"{o}dinger equation becomes  \begin{eqnarray}\label{Schal}[i\hbar (\frac{d}{dt}+A(t))\psi(t)]_{g(t)}=[\tilde{H}\psi(t)]_{g(t)}\nonumber\\\equiv [i\hbar (\frac{d}{dt}+A(t))\psi(t)=\tilde{H}\psi(t)]_{g(t)}.\end{eqnarray}

          The wave function for a free particle with energy $E$ and $\alpha(t)=0$ for all $t$   is $\psi(t)=e^{-iEt/\hbar}$ (normalization suppressed).  The Hamiltonian is a kinetic energy operator. Replacement of $\psi(t)$ in Eq. \ref{Schal} with $e^{-iEt/\hbar}$ gives\begin{equation}\label{Schi}[(E+i\hbar A(t))e^{-iEt/\hbar}=\tilde{H} e^{-iEt/\hbar}]_{g(t)}.\end{equation} This equation shows that use of the wave function $[e^{-iEt/\hbar}]_{g(t)},$ in the Schr\"{o}dinger equation with $\alpha(t)$ varying,  adds an imaginary time dependent  term to the energy.

           \subsection{Light} \label{L}

           The presence of $\alpha$ also affects the energy of light as it moves along a path, $p$ in $\mathbb{G}^{\alpha}$.  Let $x=\mathbf{x},t$ be a reference point, such as the location of an observer.  Light visible  at $x$ is emitted from events  on the past light cone from $x.$ Let $\lambda(y)_{g(y)}$ be the wave length of light  emitted by a source at time $s<t$ and  space location $\mathbf{y}$ where $y=\mathbf{y},s$ is on the past light cone. The path $p$ taken by the light as it moves from $y$ to $x$ is a geodesic where $p(s)=\mathbf{p(s)},s=y$ and $p(t)=\mathbf{p(t)},t=x$.

           The wavelength of the light arriving at $x$ is the same numerical quantity in the real number structure,  $\bar{R}^{g(p(t)))}_{p(t)}$ as $\lambda(p(s))_{g(p(s))}$ is in $\bar{R}^{g(p(s))}_{p(s)}.$ Parallel transport of the wavelength at $y$ to $x$ gives\begin{equation}\label{lamppt}\lambda'(p(t))_{g(p(t))}=[e^{-\alpha(p(t)) +\alpha(p(s))} \lambda(p(s))]_{g(p(t))}\end{equation} as the wavelength of the light arriving at $x$.

          Comparison of this wavelength with the wavelength, $\lambda(p(t))_{g(p(t))}$ of light emitted at $x$ from the same source type  as that at $y$ shows whether the wavelength has expanded or contracted  in moving from $y$ to $x$.  The wavelength has expanded or contracted if $\alpha(p(s))-\alpha(p(t))$ is greater than or less than $0.$  These changes correspond to a  respective red shift or a blue shift.

          The description of the physical property, blue or red shift of light by mathematical parallel  transport of numerical quantities   may seem strange.  It is not.  One must keep in mind that the  mathematics used to describe  physical properties of any system, with or without rest mass,  is that at the location of the system.  As the system moves along a world line the mathematics colocated with the system changes along with the motion.  The amount of change is determined by the meaning field, $\alpha.$

\begin{figure}[h]
\centering\includegraphics[width=\columnwidth]{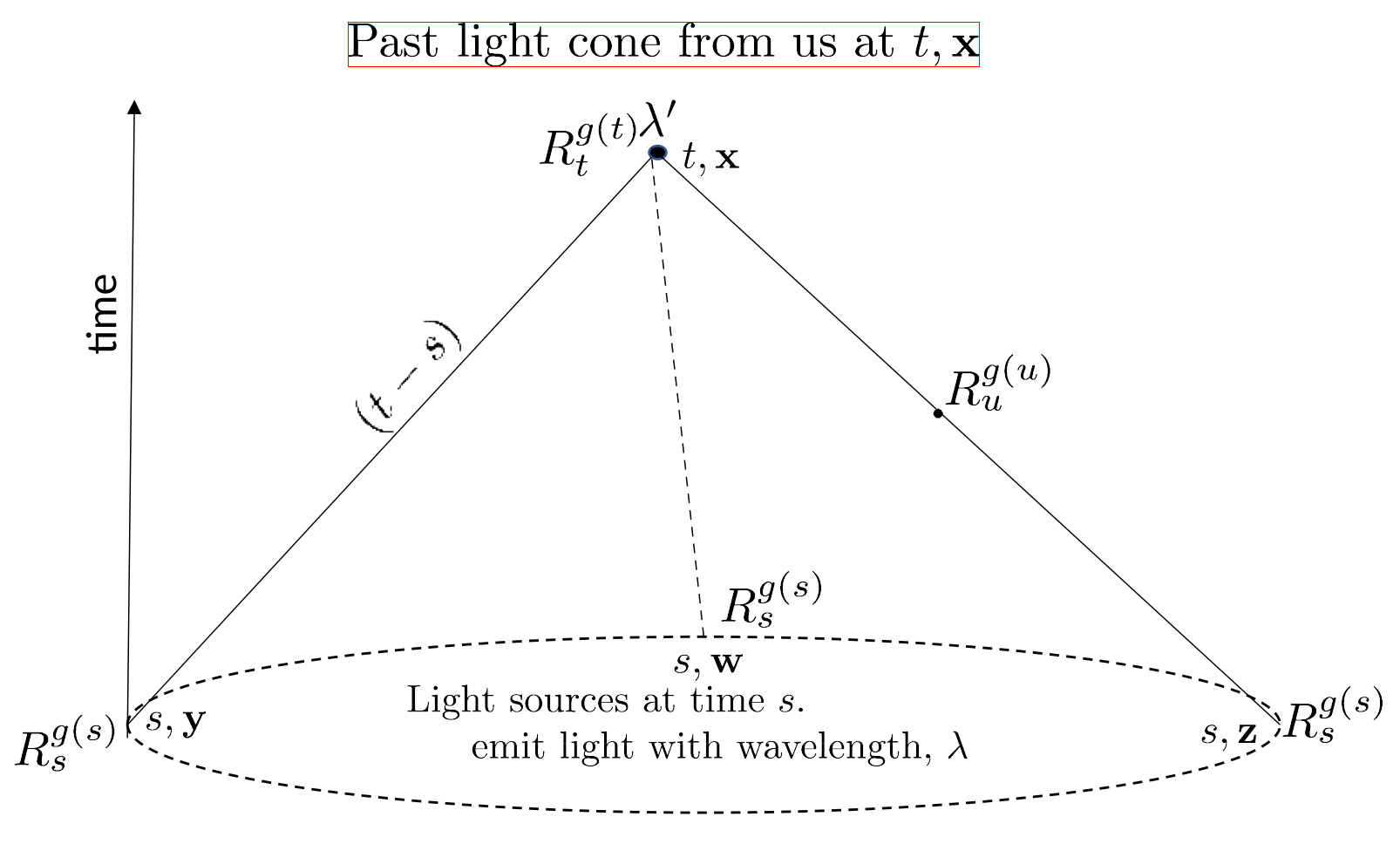}
\caption{Representation of light cone paths  from a source at time $s$ to observers at $x=\mathbf{x},t.$ Real number structures  associated with the light as it moves along the path are shown for three locations, the source at $s$, midway at $u,$ and  reception at $t.$ The wavelength $\lambda$ of the light  at the source becomes the wavelength $\lambda'$ at reception at $t.$}
\label{LSPC3} 
\end{figure}

            Figure \ref{LSPC3} illustrates this association of light with $\alpha$ dependent real number structure along a path. The illustration is for a simple case with past light cone paths.  The  field $\alpha$ is restricted to be time dependent and independent of space locations.  Three specific path location real number structure associations are shown.   These are the source locationat $s$,  midway  at $u,$ and reception by us at $t$ are shown. The wavelength values at $s$ and $t$ are shown as $\lambda$ and $\lambda'.$

           \subsection{Gauge theories}

             The gradient $\vec{A}$ field affects gauge theories.  This is a result of its effect on derivatives in Lagrangians.  The Dirac Lagrangian provides an example.

             The Dirac Lagrangian density in the presence of the $\alpha$ field is given by,\begin{equation}\label{LDy}\begin{array}{l} LD(\psi(y)= \bar{\psi}(y)i\gamma^{\mu}D_{\mu,y}\psi(y)-m\bar{\psi}(y)\psi(y)\\\\\hspace{1cm}=\bar{\psi}(y) i\gamma^{\mu}(\partial_{\mu,y}+dA_{\mu}(y))\psi-m\bar{\psi}(y)\psi(y).\end{array}\end{equation} A coupling constant, $d$  has been added to give the strength of the interaction of the $\vec{A}$  field with $\psi.$

             The Dirac action is given by \begin{equation}\label{DAx}(SD)_{g(x)} =[e^{-\alpha(x)}\int e^{\alpha(y)}(\bar{\psi}i\gamma^{\mu}D_{\mu,y}\psi -m\psi)dy]_{g(x)}.\end{equation}This action is obtained by parallel transport of the Dirac Lagrangian density at points, $y$ to a reference point, $x$. The action integral is defined because, after transport the integrands  are all scalar quantities in a single number structure, $\bar{C}^{g(x)}_{x}.$

              Minimization of the action with respect to $\bar{\psi}$ gives the Dirac equation of motion. One  obtains \begin{equation} \label{ADx} \delta S=\int e^{\alpha(y)}(\delta\bar{\psi}i\gamma^{\mu}D_{\mu,y}\psi-m\delta\bar{\psi}(y)\psi(y)) dy=0.\end{equation} Since $e^{\alpha(y)}$ and $\delta\bar{\psi(y)}$ are common factors for the integrand one obtains,\begin{equation}\label{DEM}i\gamma^{\mu} D_{\mu,y} \psi-m\psi=i\gamma^{\mu}(\partial_{\mu,y}+dA_{\mu}(y))\psi -m\psi(y)=0\end{equation} as the Dirac equation of motion. The effect of the $\alpha$ field appears in the presence of the interaction between the $\vec{A}$ field and $\psi.$

         \section{Contact with experiment}\label{CE}
         \subsection{No experiment support for variable $\alpha$}

          At this point the fact must be faced that no experiment done to date has shown the presence of a varying $\alpha$ field.  All experiments support the special case in which $\alpha(x)=0$ everywhere. No experiment supports the nonconservation of energy for free particle motion\footnote{This excludes the effect of gravity.} or  the $\alpha$ dependent position expectation values for quantum mechanical particles.

           \subsubsection{Limitation to our local region of the universe}

          There is an important caveat. The experiments supporting no effect from $\alpha$ were done in the local region, $\mathbb{R}$, of the universe. Measurements\footnote{ Experiments consist of two steps, preparation of a system in some state followed by measurement of some property of the system so prepared.  Measurements are more general in that they include determining properties of local or far away systems as they are.  No preparation step is needed. These include the sun and far away galaxies.} of properties of local systems also show no effect of $\alpha.$

          As noted before the local  space time region, $\mathbb{R}$,  includes  locations in the universe that are occupied by us.  So far this includes the surface of the earth and the international space station.  If one assumes that no effect of the $\alpha$ field will be found in regions occupiable by us now or in the future, then $\mathbb{R}$ must be expanded.

          The  size of this region is not known.  It is safe to assume that $\mathbb{R}$ is the region occupied by the solar system. If we are able to occupy regions further out and find no effect from variations in $\alpha$ then the region will have to be expanded. The ultimate size of the region  is limited by  the transit time of information from distant locations and the ability to distinguish information from noise in signals emitted by distant intelligent beings. The only important requirement on the present size of our local space time region is that it is a very small fraction of the size of the universe.\footnote{An estimate of the size of the region in which we can discover the existence of other potential civilizations  is restricted to locations within $1000$ light years from us \cite{Smith}.}

          \subsection{Restrictions on $\alpha$ variation in $\mathbb{R}$}

          The experimental indistinguishability  condition puts very stringent conditions on the variability of $\alpha$ over $\mathbb{R}.$  The deviation of $\alpha(y)$  from $0$ must be less than the uncertainty associated with any experiment.

          A reasonable estimate of the minimal uncertainty of any relevant experiment or measurement  can be had by looking  at the uncertainties associated with the various physical constants. A table of the physical constants \cite{Wikiconst} shows accuracy to  $10$ to $12$ significant figures.  This means that the  maximum deviation of $\alpha$ away  from $0$ must be less than the fractional uncertainty associated with the constants.   The deviation of $\alpha$ away from $0$ must be less than $10^{-10}$ to $10^{-12}.$

          This is a very stringent requirement on the variation of $\alpha(y)$ in $\mathbb{R}.$. However, as noted, this requirement holds only for local physical systems.  It need not hold for systems in regions outside $\mathbb{R}.$  Also the fact that variations in $\alpha$ are too small to be seen locally does not mean that variations will not be observed over cosmological distances.

         The lack of experimental support for a variable $\alpha$ in $\mathbb{R}$ means that  physical theories that include the effect of the $\alpha$ field in $\mathbb{R}$ and those that ignore the effect cannot be distinguished experimentally. The usual physical theories that ignore the effect of $\alpha$ ($\alpha(x)=0$ everywhere in $\mathbb{R}$)  can be used for the physics of systems in $\mathbb{R}.$

          \subsection{Effect of $\alpha$ on computer outputs and experimental outomes}

          An essential component of theory support or refutation is the comparison of  computer outputs, as theory predictions, with  outcomes of experiments. Computer outputs and experimental outcomes are strings of symbols, usually as digit strings.  As such they represent numbers.

          These digit strings as numbers, by themselves, have no intrinsic meaning or value. Any value  is possible.  The meaning or value of the symbol strings is determined by the $\alpha$ factor of the number structure colocated with the computer outcome or experimental outcome. If $w$ represents the computer output or experimental outcome at location $y$, the value of $w$ as a number in $\bar{S}^{g(y)},$ is $e^{-\alpha(y)}w.$ An equivalent representation of $w$ in $\bar{S}^{g(y)}$ is the number $[e^{-\alpha(y)}w]_{g(y)}.$

          A consequence of the location dependence of number values is that the values depend on where the computation or experiment are implemented.  It follows that one could have the situation that experimental support or refutation of theory depends on the locations of the computation and experiment.

          The fact that experiments, measurements, and comptations are necessarily inplemented in the restricted region, $\mathbb{R}$ and the restrictions on $\alpha$ in $\mathbb{R}$ make any such location effect unobservable. Uncertainties in experiments and possibly computations will swamp these effects.  They can be ignored.

           \section{$\mathbb{G}^{\alpha}$ and large scale properties of the universe}\label{Gmu}

             \subsection{Introduction}
            So far the domain of $\mathbb{G}^{\alpha}$ under  discussion  has been  the small region $\mathbb{R}$ of the universe.  It is clear  from the description of $\mathbb{G}^{\alpha}$ that it can represent a model of  a universe. It will be seen thaat it can model some  aspects of the real cosmological universe.

            Locations  $y$ used to describe local physics in $\mathbb{R}$ correspond to cosmological locations $z=y+x$ where $x=t,\mathbf{x}$  represents our location in the universe. Here $t$ is about $14$ billion years and $\mathbf{x}$ is our spatial location in the universe.\footnote{The location $x$ is our collective location as observers in the universe.  Strictly this is not correct as each of us has a different location in the universe. Observer $n$ is located at $x_{n}$ where $x_{n}=z_{n}+w.$  This distinction will be ignored  in $\mathbb{G}^{\alpha}$.}

            The background arena for the geometries can be considered to consist of local mathematical structures at each location in the  flat  background.  For each point $y$ in the universe there is associated local mathematics $M^{g(y)}.$   The locality of mathematics also means that the   mathematics available to an observer is that at the observers position in a world line.  The local mathematics associated with any event in the past light cone of an observer is that at the location of the event.

           This representation with  separate mathematical structures at each space time location in the background, is quite cumbersome. An equivalent representation, which is less cumbersome, is to consider $\alpha$ as a scalar field on the flat Euclidean background\footnote{To keep things simple spherical and hyperbolic backgrounds are not described.}. For each location, $y$, $\alpha(y)$  determines the relation between number and number value, vector and vector value, etc.. The explicit relation for numbers and vectors is as follows:  If $r$  and $v$ are a number and a vector at $y,$ their corresponding value or meaning is $e^{-\alpha(y)}r$ and $e^{-\alpha(y)}v.$

           If $\alpha(y)=0$, then $r$ and $v$ represent a number and vector, and their values.  Number and vector are conflated with number value and vector value.

          \subsection{Expansion of the universe in $\mathbb{G}^{\alpha}$}

           As was noted before the form of the  metric tensor,  $g_{\mu,\mu}(y)=e^{\alpha(y)} \eta_{\mu,\mu},$ means that $\mathbb{G}^{\alpha}$ cannot describe model universes with  metric tensors that have nondiagonal components, as in general relativity.  Even so there are properties of the real universe that can be described in an $\alpha$ dependent geometry.

           An example of these  properties is the time dependent  space expansion of the  universe. At any time $s<t$ the rate of space expansion is independent of space location.   It is also independent of a space dependent  mass distribution in the universe.

           This suggests that  an $\alpha$ field that depends on time and is independent of space location may be able to model the space expansion of a universe that is spatially homogenous and isotropic for free falling observers.   For each $y=s,\textbf{y}$ $\alpha(y)=\alpha(s)$.

           For this universe with a flat space background the line element is given by the FLRW equation, \begin{equation}\label{FLRW}c^{2}d\tau^{2}= c^{2}ds^{2}-a^{2}(s)d\mathbf{y}^{2}=c^{2}ds^{2}-a^{2}(s)(dr^{2}+r^{2}d\Omega) .\end{equation} The right hand term uses spherical polar coordinates for the space part where  $$d\Omega=d\theta^{2}+sin^{2}\theta d\phi^{2}.$$   The dimensionless  coefficient $a(s)$ governs the time dependent expansion or contraction of the universe.  The velocity of light, $c$ is included to be consistent with the rest of this work.

           These two equations are for values of numerical quantities.  The equations for numerical quantities are obtained by adding the subscript $g(s)$ to the terms of the equations.

            The time depenndent scale factor  $g(s)=e^{\alpha(s)}$ is the  scale factor for mathematical structures, $M^{g(s)}$ at different times, $s$.\footnote{$M^{g(s)}$ is global in space and local in time.} Parallel transport of the value of a numerical quantity at $s$ to time $t$ is implemented by the connections $C_{g}(t,s).$

            The wavelength of light provides a good example.  If $\lambda_{g(s)}$ represents the wavelength of light emitted from an event at time $s$, then
            \begin{equation}\label{ldbap}[\lambda']_{g(t)}=C_{g}(t,s)\lambda_{g(s)}=[e^{-\alpha(t) +\alpha(s)} \lambda]_{g(t)}\end{equation} represents the wavelength of the light at a  later time $t$.

            The relation between the wavelengths and the scale factors \cite{Weinberg} at time $t$ is given by, \begin{equation}\label{latlas}[\frac{\lambda'} {\lambda}]_{g(t)}= [\frac{a(t)} {a(s)}]_{g(t)}=\frac{a(t)_{g(t)}}{a(s)_{g(t)}}.\end{equation}The implied righthand division is an operation in $\bar{R}^{g(t)}.$

            The number $a(s)_{g(t)}$ in $\bar{R}^{g(t)}$ has the same value  as the number $a(s)_{g(s)}$ has in $\bar{R}^{g(s)}.$ However  $a(s)_{g(t)}$  and $a(s)_{g(s)}$ are different numbers. Also the number $\lambda_{g(t)}=\lambda(s)_{g(t)}$ has the same value in $\bar{R}^{g(t)}$ as $\lambda(s)_{g(s)}$ has in $\bar{R}^{g(s)}.$ They are also different numbers. Measurement of the wavelength of light at time $t$ from the same source as that at $s$ is assumed to give light with wavelength $\lambda_{g(t)}.$

            From Eq. \ref{ldbap} one sees from Eq. \ref{latlas} that \begin{equation} \label{east}[\frac{a(t)} {a(s)}]_{g(t)}=[e^{-\alpha(t)+\alpha(s)}]_{g(t)}. \end{equation} Normalizing the scale factor to unity, $a(t)=1,$ for the present time, $t$  in Eq. \ref{east} gives\begin{equation}\label{aalphs} a(s)=e^{-\alpha(s)}\end{equation} as the relation between the scale factor $a$ and the value field $\alpha$. If $\alpha(s)$ \emph{decreases} as time increases, then the scale factor value, $a(s),$ \emph{increases} as the time, $s$ increases.

            Eq. \ref{ldbap} shows that  the value field factors for different $s$ are all numerical quantities, $[e^{-\alpha(s)}]_{g(t)}$ in $\bar{R}^{g(t)}.$ It follows that all the usual mathematical operations on the value field are defined within $\bar{R}^{g(t)}$ and are defined.  For example the derivative, \begin{equation} \label{sdea}\frac{d}{ds}[e^{-\alpha(s)}]_{g(t)}=[-A(s)e^{-\alpha(s)}]_{g(t)} \end{equation} is defined.\footnote{Note that $D_{s}[e^{-\alpha(s)}]_{g(s)}$  equals $0.$} Here \begin{equation}\label{Asfr}A(s)= \frac{d\alpha(s)} {ds}.\end{equation}

           \subsubsection{Hubble expansion}

           Unless specifically noted, from here on the equations  are all in terms of \emph{values} of numerical and vectorial quantities at time $t$.  The equations in terms of numerical or vectorial quantities  are obtained by addition of the subscript $g(t)$.

           The relation between $a(s)$ and the Hubble parameter, $H(s)$ is given by\cite{Weinberg}
           \begin{equation}\label{Hs}H(s)=\frac{\dot{a}(s)}{a(s)}\end{equation} where $\dot{a}(s)$ is the time derivative of $a(s)$. Replacement of $a(s)$ by $e^{-\alpha(s)}$, Eq. \ref{aalphs}, gives \begin{equation}\label{Hsal}H(s)= e^{\alpha(s)}\frac{de^{-\alpha(s)}}{ds} =-A(s).\end{equation} Here \begin{equation}\label{Asda} A(s)=\frac{d\alpha(s)}{d(s)}
            \end{equation}is negative.

           The Hubble constant, $H_{0}=H(t)$, is the Hubble parameter at the present time, $t$.
           Fom Eq. \ref{Hsal} one has \begin{equation}\label{cAc}-A(t)=H_{0}.\end{equation}  For nearby sources  $H(s)\approx H_{0}$. This shows that $A(s)\approx A(t)$ for these sources.

           The experimental value of $H_{0}$ appears to depend on how it is determined. Resolution of this conflict is a topic of much research \cite{Freeman}.  Here the exact value is not important so $H_{0}$ is a rough average of the empirical values \cite{WikiHub} with a value of $70$ km/sec/megaparsec. A megaparsec is $3.26$ million light years.

           From Eq. \ref{cAc} one sees that \begin{eqnarray}\label{AtH0}-A(t)=H_{0}=70\mbox{ km/sec/megaparsec}\nonumber\\ =7.16\times 10^{-11}/year=2.3\times 10^{-18}/sec.\end{eqnarray} This value of $A(t)$ supports the assertion made earlier that the effect of the field $\alpha$ on temporal aspects of  physics in the local region, $\mathbb{R}$,\footnote{The effect of $A(t)$ in the examples described in section \ref{LPG} is limited to the time dependent Schr\"{o}dinger equation and the time component of the Dirac Lagrangian.}  is far too small to be detected in experiments. The red  shift of light from a source $300,000$ km from us at $x$ would have to be measured to an accurcy of $18$ significant figures to be detectable.  The accuracy of a measurement for light from the sun at $8$ light minutes distant would have to be to at least  $15$ significant figures.

            The examples show that local physics with a time dependent $\alpha$ whose gradient satisfies Eq. \ref{AtH0} is experimentally indistinguishable from the usual physics with $\alpha(t)=0$. For this physics  the time dependent change in energy of a free particle moving along a geodesic as in Eq. \ref{Emps} disappears. Energy is conserved. The wavelength shift of light from local sources, Eq. \ref{lamppt}, is too small to be observed.

    \subsubsection{The red  shift parameter in $\mathbb{G}^{\alpha}$}

    The amount of red (or  blue) shift of the wavelength values of light from distant sources is represented by the shift parameter, $z$, given by    \begin{equation}\label{zm1}z= \frac{\lambda'} {\lambda}-1=e^{\alpha(s)- \alpha(t)}-1. \end{equation}A positive exponent gives a red shift; a negative exponent gives a blue shift.

            For nearby sources $\alpha(s)\approx \alpha(t)$. Expansion of the exponential to first order in small quantities gives \begin{equation}\label{zalp}
            z\approx \alpha(s)-\alpha(t).\end{equation}Writing $\alpha(s)=\alpha(t-(t-s))$ and using a Taylor expansion gives  \begin{equation}\label{zts} z\approx\alpha(t)-(t-s) A(t)-\alpha(t)=-(t-s)A(t)=(t-s)H(t)\end{equation} in agreement with \cite{Weinberg}. Eq. \ref{Hsal} has been used here.

            The  $t$ time derivative of $z$ in Eq. \ref{zts} gives\begin{equation}\label{dzdt} \frac{dz}{dt} \approx H(t)+(t-s)\frac{dH(t)}{dt}\approx H(t).\end{equation}Here $dH(t)/dt=-dA(t)/dt$ is set equal to zero. Use of Eq. \ref{cAc} gives\begin{equation}\label{apprH}\frac{dz}{dt}\approx H(t)=H_{0}= -A(t). \end{equation}
            \subsection{The Friedmann equations}

            There are two Friedmann equations \cite{Friedmann,Weinberg} that are solutions of the Einstein equations for an isotropic homogenous  flat ($K=0$) universe. They are \begin{equation} \label{Fr1}\frac{\dot{a}^{2}}{a^{2}}=\frac{8\pi G\rho}{3}\end{equation} and \begin{equation}\label{Fr2}\frac{\ddot{a}}{a}=-\frac{4\pi G}{3}(\rho+\frac{3p} {c^{2}}).\end{equation} In these equations $a$ is the time dependent spatial scaling factor in the FLRW equation, Eq. \ref{FLRW}, $G$ is Newton's Gravitational constant, $\rho$ is the energy density, $p$ is the pressure,  and $c$ is the velocity of light.

            These equations do not include the effect of the presence of the cosmological constant, $\Lambda$ .  This can be taken into account by addition of the term $\frac{\Lambda c^{2}}{3}$ to  the right side of both \cite{FriedWiki} of the equations. Here $\Lambda$ is  in terms of inverse length squared.

            The relation between the gradient of $\alpha$ and the Friedmann equations is obtained from Eq. \ref{aalphs} as\begin{equation}\label{FrA1}\frac{\dot{a} (s)^{2}}{a(s)^{2}} =A(s)^{2} =\frac{8\pi G\rho(s)}{3}\end{equation} and \begin{equation}\label{FrA2} \frac{\ddot{a}(s)}{a(s)}=-\dot{A}(s)+ A(s)^{2}=-\frac{4\pi G}{3}(\rho(s)+\frac{3p(s)} {c^{2}}).\end{equation}  Combining Eqs. \ref{FrA1} and \ref{FrA2} to remove $A(s)^{2}$ gives\footnote{These equations are for values of numerical physical quantities. The equations for numerical physical quantities are obtained by addition of the subscript $g(s)$ to all the termss.}\begin{equation} \label{dtAs}\dot{A}(s)=4\pi G(\rho(s)+ p(s)).\end{equation}  This equation is unaffected by the presence of the cosmological constant term $\frac{\Lambda c^{2}}{3}$  added to the righthand terms \cite{FriedWiki} in Eqs. \ref{FrA1} and \ref{FrA2}.

            This equation shows an acceleration  of space expansion if $\dot{A}(s)>0$  or $\rho(s)+p(s)>0.$  It shows a deceleration of space expansion or space contraction if $\dot{A}(s)<0$  or $\rho(s)+p(s)<0.$  The expansion or contraction rate is constant  if  $\dot{A}(s)=0$ or $\rho(s)=-p(s).$  The sign of $-A(s)$ determines whether space is expanding or contracting. $A(s)=0$ describes a static universe.

            For a flat universe, $K=0$, the energy  density, $\rho_{t},$  at the present time,$t$ is equal to the present critical energy density \cite{Weinberg} $\rho_{t,c}$. From Eq. \ref{FrA1} one obtains\begin{equation}\label{rtc}\rho_{t,c}= \frac{\dot{a}(t)^{2}} {a(t)^{2}}\frac{3}{8\pi G}=\frac{3H_{0}^{2}}{8\pi G}=\frac{3A(t)^{2}}{8\pi G}.\end{equation}

            The total energy density is made  up of three components, the matter energy density, $\rho_{m}$, the radiation energy density, $\rho_{r}$ and the vacuum energy  density, $\rho_{v}.$ The average energy density at time $s$, as a weighted sum of the component densities,  is given by \cite{Weinberg} \begin{eqnarray}\label{ohrt} \rho(s)= \Omega_{m}\rho_{m}(s)+\Omega_{r}\rho_{r}(s)+\Omega_{v}\rho_{v}=\nonumber\hspace{1cm}\\ \frac{3A(t)^{2}}{8\pi G}(\Omega_{m} \frac{a(t)^{3}} {a(s)^{3}} +\Omega_{r}\frac{a(t)^{4}}{a(s)^{4}}+\Omega_{v}) .\end{eqnarray}The coefficients $\Omega_{m}, \Omega_{r}, \Omega_{v}$ are the fractions of the component matter, radiation, and vacuum densities at the present time, $t,$ in the universe. Use of Eq. \ref{aalphs} to replace $a(s)$ and $a(t)$ with $e^{-\alpha(s)}$ and $e^{-\alpha(t)}$ gives\begin{equation}\label{AGHa} \rho(s)=\frac{3A(t)^{2}}{8\pi G}(\Omega_{m}(e^{-\alpha(t)+ \alpha(s)})^{3}+\Omega_{r} (e^{-\alpha(t)+ \alpha(s)})^{4} +\Omega_{v}).\end{equation}

            The time dependences of the  densities, $\rho_{m}(s)$ or $\rho_{r}(s)$ for universes consisting of just matter or just radiation are represented by factors multiplying the $\Omega$ coefficients in Eq. \ref{ohrt}. The resulting expressions for the densities are used in Eq. \ref{FrA1} to obtain the resulting time dependences. Combining these dependencies  with the relation between $a(s)$ and the scaling factor $\alpha(s)$ in Eq. \ref{aalphs}  gives \cite{Weinberg} \begin{equation} \label{rhom}e^{-\alpha(s)}=a(s)\propto s^{2/3}\end{equation}for nonrelativistic matter and \begin{equation}\label{rhor}e^{-\alpha(s)}=a(s)\propto s^{1/2}\end{equation} for radiation.

             For the vacuum energy one replaces the time dependent density by a constant vacuum emergy density, $\rho_{V}$ and solves Eq. \ref{FrA1}. One obtains, with Eq. \ref{aalphs},\begin{equation}\label{rhoV}e^{-\alpha(s)}=a(s)\propto e^{(\frac{8\pi G\rho_{V}}{3})^{1/2}s}.\end{equation}

             An equivalent representation of this  equation is\begin{equation}\label{rhoVq} e^{-\alpha(s)}= e^{-(\frac{8\pi G\rho_{V}}{3})^{1/2}s+d}\end{equation}where $d$ is an arbitrary constant. The constant $d$ can be determined by setting the value of the scale factor, $a$ equal to $1$ at the present time, $t$.This gives $\alpha(t)=0.$ From Eq. \ref{aalphs} one obtains \begin{equation}\label{esd}e^{d}=e^{(\frac{8\pi G\rho_{V}}{3}) ^{1/2}t}.\end{equation}Equating exponents in Eq. \ref{rhoVq} gives \begin{equation} \label{alpsf}\alpha(s)=(\frac{8\pi G\rho_{V}}{3})^{1/2}(t-s)=H_{0}(t-s).\end{equation}Eq. \ref{rtc} has been used here.

             This equation is valid only for systems that are not too distant with small red shifts. For these systems the expansion rate of the universe is close to the present day value of the Hubble parameter, $H_{0}.$.  It does not include early times close  to the big bang or  space acceleration due to dark energy.

             \subsection{$\alpha$ representation of dark energy and  big bang space  expansion}

             The  value field, $\alpha$, dependence on time is sufficiently flexible so that it can include the big bang at time $s=0$ and the  space expansion acceleration caused by dark energy.  For the big bang the FLRW scale factor $a(s)$ is set equal to $0$ for $s=0.$  From Eq. \ref{aalphs} this is represented by setting $\alpha(0)=\infty.$   For times $s$ in which the universe expansion is accelerating, $\dot{A}(s)<0.$ As before, $A(s)$ is the gradient of $\alpha(s).$

\begin{figure}[h]
\centering\includegraphics[width=\columnwidth]{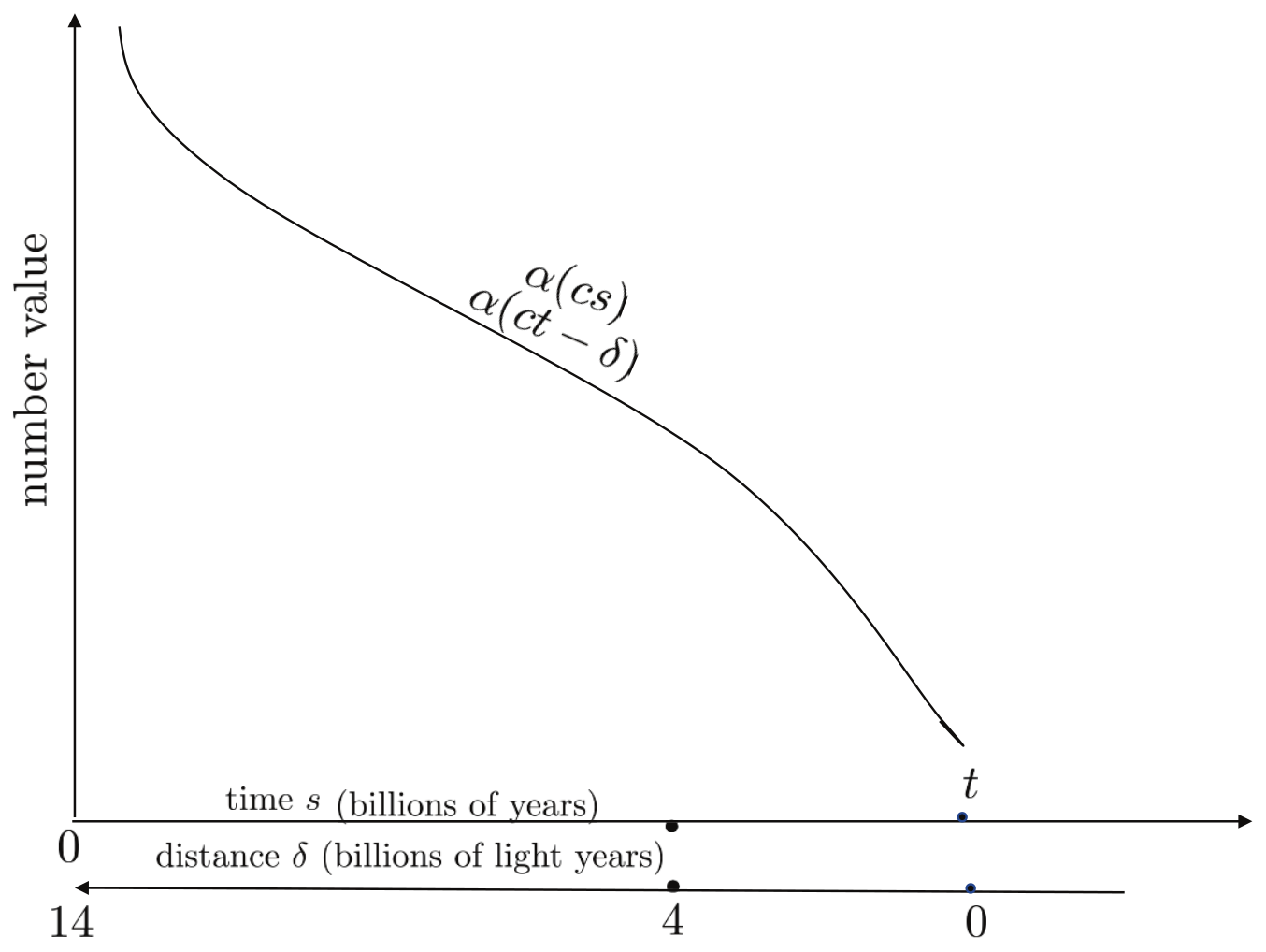}
\caption{Shows a possible dependence of the value field, $\alpha$ on cosmological time and distance.  The time, $s$ ranges from $0$, the time of the big bang, to the present time, $t$. The distance extends from $0$ at our location back to the big bang at $14$ billion light years. The upturn of the curve for small $s$ values is followed by a region with a constant negative slope. This is followed by a downturn starting about $4$ billion years ago.}\label{LSPC1} 
\end{figure}

             The time of onset of dark energy space expansion  is set at $4$ billion light years ago \cite{WikiAc,Frieman},  at $s\approx 10$ billion years. Use of this condition and the fact that $\alpha(0)=\infty$ leads to the functional form of the time dependence of $\alpha$   shown in Figure \ref{LSPC1}. The region of the curve with constant negative slope represents the Hubble expansion. The increase in negative slope  $4$ billion years ago \cite{DkEnb} represents the onset of accelerated space expansion ascribed to dark energy \cite{Peebles}. 

        The rapid upward approach of $\alpha(cs)$ to $\infty$ as $s\rightarrow 0$ represents the contraction of  space in $\mathbb{G}^{\alpha}$ to a point.  This  is a  representation of the big bang. From Eq. \ref{aalphs} one sees that the scale parameter, $a(s)\rightarrow 0$ and $\alpha(s)\rightarrow \infty$ as $s\rightarrow 0.$

        It is to be emphasized that Fig. \ref{LSPC1} is a free hand drawing. It is drawn to represent the information available to date.  No calculations are involved. It is hoped that, in the future, more information will become available to create a more accurate figure.

        \section{Discussion}

        \subsection{Summary}

        This work is based on the distinction between number and number value or meaning.  This and the no information at a distance principle, or freedom of choice at a distance, leads to the replacement of the usual global mathematics by local mathematics. Local mathematics consists of those mathematical structures that include numbers in their axiomatic description.

        Local mathematical structure dependence at space time locations is accounted for by   the presence of a space time dependent scaling or meaning factor.  For each mathemmatical structure at each location the scaling factor assigns values to each structure component.

        Example for real numbers: Let $\bar{R}^{g(y)}$  be the real number structure at location $y$ with scale factor $g(y).$  Real numbers and real number variables in $\bar{R}^{g(y)}$  have the form $\pi_{g(y)}$ and $r_{g(y)}$.  Here $\pi_{g(y)}$ is the number with value $\pi$ and $r_{g(y)}$ is the number variable with number variable value $r$.  Scaling of number structures is normalized by number value and number being identified in  structures where $g(y)=1.$

        Relations between values of numbers in structures at different locations are defined by means of a number preserving value changing connection.  If $\pi_{g(y)}$ is a number in $\bar{R}^{g(y)}$, then this same number is represented by $$C_{g}(x,y)\pi_{g(y)}= [\frac{g(y)}{g(x)}\pi ]_{g(x)}$$ in $\bar{R}^{g(x)}.$

        From now on a positive number valued field, $\alpha$ where $e^{\alpha(y)}=g(y)$ will replace $g(y)$ in expressions. The nomenclature, $g(y)$ will be retained for most superscripts and subscripts.

        The presence of the $\alpha$ field and local mathematics affects description of system properties in theoretical physics. This is a result of  the implied arithmetic combinations between numbers and vectors at different locations  in  the definitions of integrals and derivatives over space, time, and space time.

        This is fixed by the use of connections to parallel transport arithmetic elements to a common location for combination. For a function $f(y)$ the resulting integral is
        \begin{equation}\label{Cint}\int\prod_{y}C_{g}(x,y)[f(y)_{g(y)}dy]_{g(y)}=[e^{-\alpha(x)} \int e^{\alpha(y)dy}f(y)dy]_{g(x)}.\end{equation} The resulting derivative is
        \begin{eqnarray}\label{Dfmuy}D_{\mu}f(y)=\frac{C_{g}(y,y+d_{\mu}y)f(y+d_{\mu}y)_{g(y)} +d_{\mu}y)-f(y)_{g(y)}} {[d_{\mu}y]_{g(y)}}\nonumber\\=[(\partial_{\mu,y}+A_{\mu} (y))f(y)]_{g(x)}.\hspace{4cm}\end{eqnarray}The limits are understood. Here $x$ is the reference  location and\begin{equation}\label{yumA}A_{\mu}(y)=\frac{d\alpha(y)} {d_{\mu}y}.\end{equation}

        Local mathematics and the space time dependent scale or value field, $\alpha$ provide a background for the description of local physics and cosmology in a geometry $\mathbb{G}^{\alpha}.$  The geometry is divided into two regions,  a local region $\mathbb{R}$ and the rest of the universe.

        The local region consists of the part of the universe occupiable by us as intelligent observers. Local physics is the physics done by us  in $\mathbb{R}.$ Cosmological properties of a model universe are described in $\mathbb{G}^{\alpha}.$

        Deviations of $\mathbb{G}^{\alpha}$ from flatness depend on the extent of variations in $\alpha.$ This is expressed by the metric tensor $[e^{-\alpha(x)+\alpha(y)}\eta]_{g(x)}$ of $\mathbb{G}^{\alpha}.$ Here $\eta$ is the space time metric tensor, $x$ is a  reference point in $\mathbb{R}$ and $y$ is any point in $\mathbb{G}^{\alpha}.$

        Examples of the effect of $\alpha$ on local physics are given these include the replacement of the derivative $\partial_{\mu,y}\psi(y)$ by $D_{\mu,y}\psi(y)=(\partial_{\mu,y} +A_{\mu}(y))\psi(y)$ in the Dirac Lagrangian density. The derivative $\frac{d\psi(t)}{dt}$ in the time dependent Schr\"{o}dinger equation is replaced with $D_{t}\psi(t)=(\frac{d}{dt} +A(t))\psi(t).$

        There is no experimental evidence for the presence of $\alpha$  or the derivative, $\vec{A}$, of $\alpha$ in local physical  theory. It follows that theories with  $\alpha$ cannot be distinguished from theories without $\alpha.$  The effect of $\alpha$ must be within experimental error for all experiments.

        It turns out that $\alpha$ and its derivative can be  used to describe values of  cosnological  properties. The resulting values are too small to have an effect on any local physical experiment.

        The value field, $\alpha$ and its derivative can be used for the values of some properties of  an isotropic, homogenous universe. The  derivative of $\alpha$ can be set equal to the Hubble constant. Also $\alpha$ can be related to the scale factor in the FLRW equation.  As a result the scale factor and its derivative in the Friedmann equations can be  replaced by $\alpha$ and its derivative. To keep the work simple the description is limited to a flat universe.  It was also noted that $\alpha$ can be used account for the big bang and accelerated expansion due to dark energy.

         The use of the $\alpha$ dependent geometries shows that that it \emph{can} represent Hubble expansion, the big bang, and the acceleration due to dark energy. It  \emph{can} also be related to the scale factor in the FLRW equations and the Friedmann equations. It has not been shown that it \emph{must} represent these cosmological properties. To show this would require showing that the $\alpha$ field that describes the scaling of local mathematics and its relation to physics and geometry  is unique. It would need to be shown that it \emph{must}  have these properties.

          \subsection{Limitations and generalizations of $\alpha$}

        There are limitations on the possible geometry $\mathbb{G}^{\alpha}.$ It cannot be a complete representation of the cosmological models,  such as the $\Lambda CDM$ model. The reason is that all $\alpha$ dependent geometries have metric tensors for which the
        nondiagonal elements are all $0$. This restriction does not apply to the diagonal elements.

        The geometries $\mathbb{G}^{\alpha}$ can be more general than those described  so far for the universe. The restriction that $\alpha$ depend on time only can be lifted so that it can depend on space as well as time. These geometries remain to be investigated.

        \subsection{Future work}

        Much remains to be done. Geometries based on space and time dependent value fields should be investigated. The relation, if any, of the geometries to quintessence \cite{ZlSt,Asen,Brand}  and other scalar fields described in the literature should be examined. It should also be seen if geometry based on complex valued meaning fields make sense.

        The fact that $\alpha$ gives meaning or value to numbers and other mathematical systems and expressions in theoretical physics suggest a relation to consciousness \cite{BenCONC}.  Meaning or value is a basic, necessary  component of consciousness. It may be that investigation of this relation gives some light to why $\alpha$ must affect local physics and cosmological preperties as shown in this work.

        \end{document}